\documentclass[a4paper,11pt]{article}
\usepackage{pos}
\usepackage{lipsum} 
\usepackage{xcolor}
\usepackage{hyperref}
\usepackage{amsmath}
\usepackage{etoolbox}
\makeatletter
\pretocmd{\@cite}{\textbf{\color{blue}}}{}{}
\makeatother
\usepackage{tikz}
\usetikzlibrary{shapes.geometric, arrows}

\tikzstyle{process} = [rectangle, minimum width=2.5cm, minimum height=1cm, text centered, draw=black, fill=orange!30]
\tikzstyle{decision} = [diamond, minimum width=2.5cm, minimum height=1cm, text centered, draw=black, fill=green!30]
\tikzstyle{arrow} = [thick,->,>=stealth]

\title{ARA-Next: a new DAQ and trigger architecture for the Askaryan Radio Array }
\ShortTitle{ARA-Next , New DAQ with complex triggers}

\author*[a]{Pawan Giri}
\author[a]{Ilya Kravchenko}
\author[b]{Patrick Allison}
\author[b]{Amy L. Connolly}
\onbehalf{for the ARA Collaboration \\{\normalsize \normalfont(a complete list of authors can be found at the end of the proceedings)}\\}

\affiliation[a]{University of Nebraska - Lincoln,\\
Lincoln, NE}

\affiliation[b]{The Ohio State University,\\
Columbus, OH}

\emailAdd{pgiri4@huskers.unl.edu}

\abstract{

The Askaryan Radio Array (ARA) experiment aims to detect ultra-high-energy cosmic neutrinos (>10 PeV) using radio detection techniques. To enhance ARA's capabilities, a new RFSoC-based DAQ, ARA-Next, is in the early stages of development. This advanced system will facilitate the creation of sophisticated triggers, including a novel multi-trigger approach, similar to  those used in collider experiments. Our approach involves crafting and implementing innovative triggers for ARA's new DAQ, such as identifying double pulses from potential in-ice neutrino interactions, utilizing templates for atmospheric cosmic ray signals, optimizing triggers for astrophysical neutrino sources, correlating special events between ARA and IceCube, and discerning anthropogenic events using directional information. These trigger designs aim to lower thresholds and enhance ARA's detector sensitivity. Overall, this upgrade will not only enhance ARA's capabilities but also contribute to the technological advancements necessary for future experiments of this nature. 

}

\FullConference{10th International Workshop on Acoustic and Radio EeV Neutrino Detection Activities (ARENA2024)\\
11-14 June 2024\\
The Kavli Institute for Cosmological Physics, Chicago, IL, USA\\}

\begin{document}
\maketitle

\section{Introduction}\label{sec1}
\begin{figure}
\centering
\includegraphics[width=0.7\linewidth]{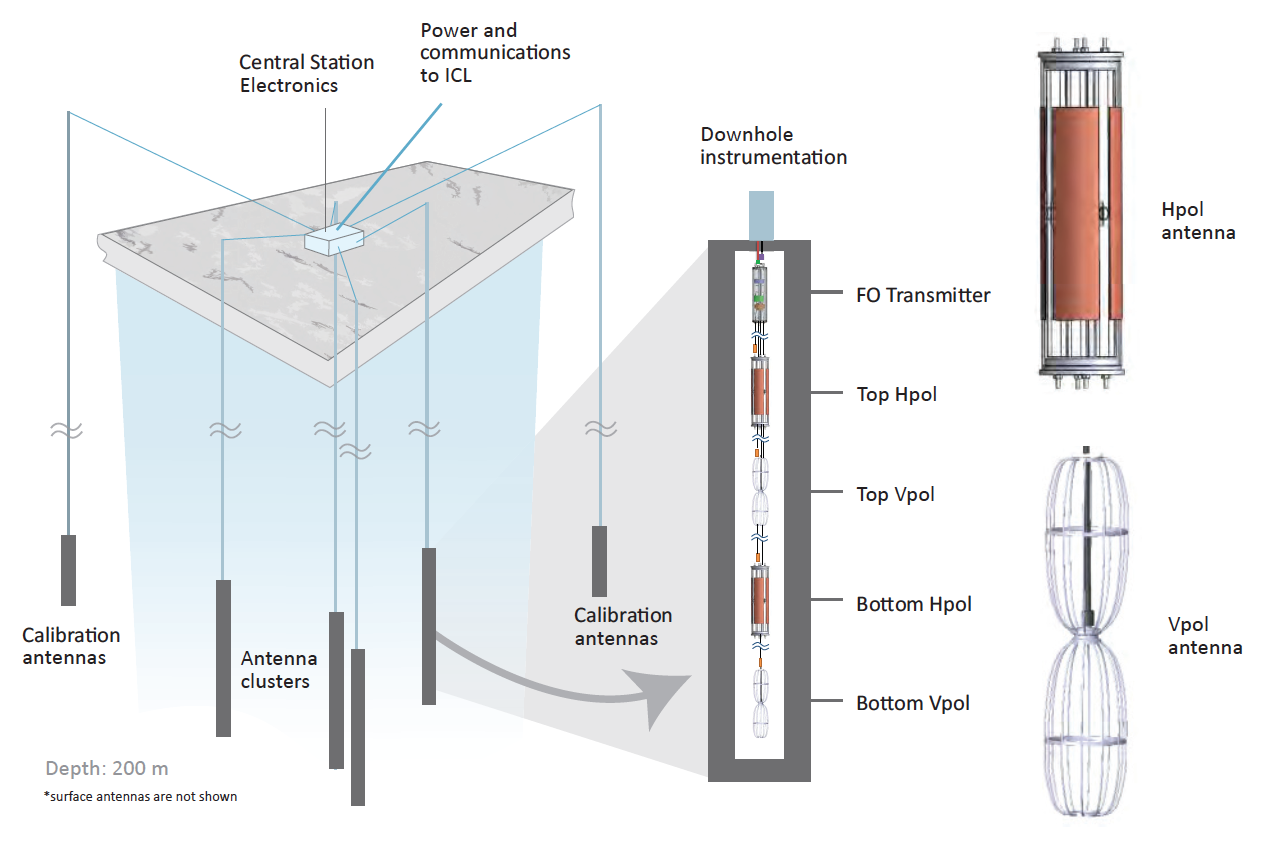}
\caption{\label{fig:Station}Schematic representation of a typical ARA station and quadslot (top-right) and birdcage(bottom-right) dipole antennas used in ARA.}
\end{figure}

Neutrinos with energies exceeding 10 PeV serve as unique messengers from distant parts of the high-energy universe. Unlike cosmic rays and gamma rays, neutrinos are neutral and weakly interacting particles, allowing them to traverse vast cosmic distances without obstruction, making it possible to trace them back to their origins \cite{ref1}.

Over the past decade, IceCube has successfully detected astrophysical neutrinos, mapped their energy spectrum up to around 10 PeV \cite{ref2,ref3}, and identified two active galaxies as sources \cite{ref4,ref5}, highlighting the critical role of neutrinos in high-energy multi-messenger astrophysics. However, IceCube’s detection capabilities are limited at higher energies due to its 1 km³ optical signature footprint. In contrast, ARA is expected to surpass IceCube’s current limits for neutrinos above $10^{17}$ eV, aligning with key predictions of astrophysical and cosmogenic neutrino models.

Our goal is to develop ARA-Next Data Acquisition (DAQ) systems for the five existing ARA stations using Radio Frequency System-on-Chip (RFSoC) technology. The ARA-Next DAQ will significantly boost ARA’s potential without expanding its footprint, allowing for new physics triggers to capture various event topologies and neutrinos from particular directions.

\section{Current ARA trigger board and trigger logic}\label{sec2}

Each standard ARA station is equipped with eight vertically polarized (Vpol) and eight horizontally polarized (Hpol) antennas, arranged in four strings as in Figure 1. Each string has four antennas, with two for each polarization. The current ARA DAQ system is built around a custom motherboard called the ARA Trigger and Readout Interface (ATRI), as illustrated in Figure 2. The voltage traces recorded by the in-ice antennas are transferred to the DAQ board, where they first pass through band-pass filters to eliminate unwanted frequencies. The signals are then split into two paths by a splitter. One path is sent to the trigger system, where triggering is performed by four Triggering DAughter boards (TDAs) \cite{ref6} , where the signals must satisfy four different stages of trigger logic - Level1 to Level4 (L1 to L4).

\begin{figure}[htbp]
\centering
\begin{minipage}[b]{0.47\linewidth}
    \centering
    \includegraphics[width=\linewidth]{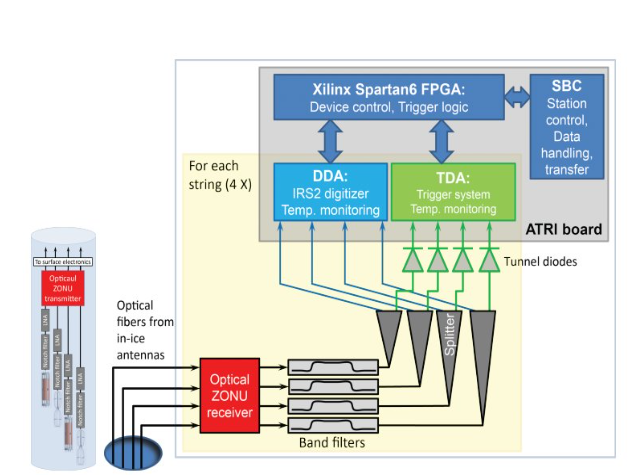}
    \caption{Current DAQ for ARA.}
    \label{DAQ}
\end{minipage}
\hfill
\begin{minipage}[b]{0.47\linewidth}
    \centering
    \includegraphics[width=\linewidth]{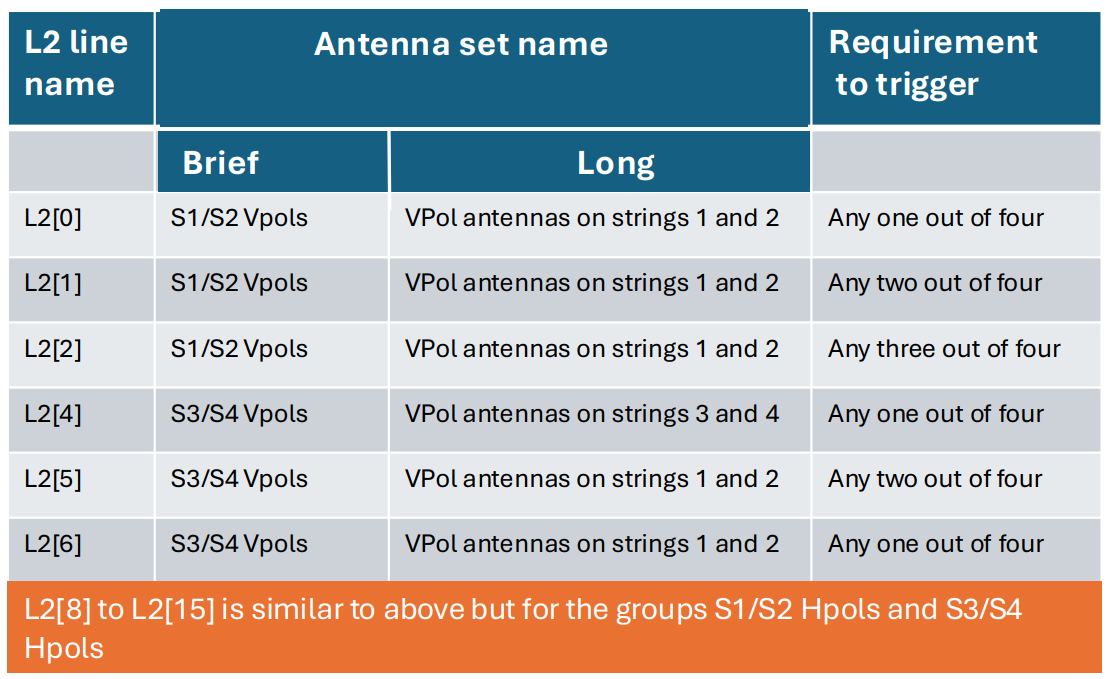} 
    \caption{L2 trigger logic table.}
    \label{fig:L2 trigger logic in brief. }
\end{minipage}
\end{figure}


An L1 trigger is a signal detected by an antenna when the threshold of the tunnel diode or power detector for that channel is exceeded. The duration of each L1 trigger varies based on the specific implementation. Essentially, the response time of the tunnel diode or power detector is around 10 nanoseconds. Therefore, an L1 trigger is considered active when the tunnel diode surpasses the threshold and stays active for at least 5 nanoseconds.
In total there are 16 L1 triggers for 16 channels namely $L1_{\textcolor{blue}{(0)}}$ to $L1_{\textcolor{blue}{(15)}}$, where $L1_{\textcolor{blue}{(0)}}-L1_{\textcolor{blue}{(7)}}$ comes from Vpols and $L1_{\textcolor{blue}{(8)}}-L1_{\textcolor{blue}{(15)}}$ comes from Hpols.

Level2 triggers, or L2 triggers, are based on combinations of L1 triggers. $L2_{\textcolor{blue}{(0)}}$ activates if any one of the four Vpols in strings 1 and 2 is triggered. $L2_{\textcolor{blue}{(1)}}$ activates if any two are triggered, $L2_{\textcolor{blue}{(2)}}$ for any three. Similarly, $L2_{\textcolor{blue}{(4)}}$ to $L2_{\textcolor{blue}{(6)}}$ apply the same logic to the Vpols in strings 3 and 4. $L2_{\textcolor{blue}{(8)}}$ to $L2_{\textcolor{blue}{(15)}}$ follow the same idea but for Hpols. The L2 trigger logic has been tabulated in Figure 3.

The Level3 (L3) trigger logic evaluates combinations of L2 triggers to determine event activation. $L3_{\textcolor{blue}{(0)}}$ is triggered when any 3 out of the 8 Vpol channels are activated within 170 ns window, with the possible L2 combinations outlined in Equation 1. Similarly, $L3_{\textcolor{blue}{(1)}}$ is triggered when any 3 out of the 8 Hpol channels are activated within the same time window, as described by the L2 combinations in Equation 2.

For Vpol channels, the L3 trigger logic is :   
\begin{equation}
L2_{\textcolor{blue}{(0)}} = L2_{\textcolor{blue}{(2)}} \text{ or } L2_{\textcolor{blue}{(6)}} \text{ or } (L2_{\textcolor{blue}{(0)}} \text{ and } L2_{\textcolor{blue}{(5)}}) ] \text{ or } (L2_{\textcolor{blue}{(1)}} \text{ and } L2_{\textcolor{blue}{(4)}}) 
\end{equation}

For Hpol channels, the L3 trigger logic is:
\begin{equation}
L3_{\textcolor{blue}{(1)}} = L2_{\textcolor{blue}{(10)}} \text{ or } L2_{\textcolor{blue}{(14)}} \text{ or } (L2_{\textcolor{blue}{(8)}} \text{ and } L2_{\textcolor{blue}{(13)}}) ] \text{ or } (L2_{\textcolor{blue}{(9)}} \text{ and } L2_{\textcolor{blue}{(12)}})
\end{equation}

The Level4 (L4) trigger represents the final stage of the decision process, activating if either $L3_{\textcolor{blue}{(0)}}$ or $L3_{\textcolor{blue}{(1)}}$ is high.


The second path is directed to digitization, managed by four Digitizing DAughter boards (DDAs). The IceRay Sampler 2 chip, a 3.2 Gs/s digitizer, stores signals in analog Switched Capacitor Arrays to conserve power. Once the L4 trigger condition is met, sampling stops, and 400-600 ns of waveform data is recorded and sent to the data collection center \cite{ref6}.

While the existing trigger system has performed well in data collection for ARA, it lacks flexibility for incorporating additional trigger logic. Given that this trigger logic has been in use for over a decade, we plan to upgrade it by replacing the ATRI with a more modern and versatile RFSoC.

\begin{figure}[htbp]
\centering
\begin{minipage}[b]{0.42\linewidth}
    \centering
    \includegraphics[width=\linewidth]{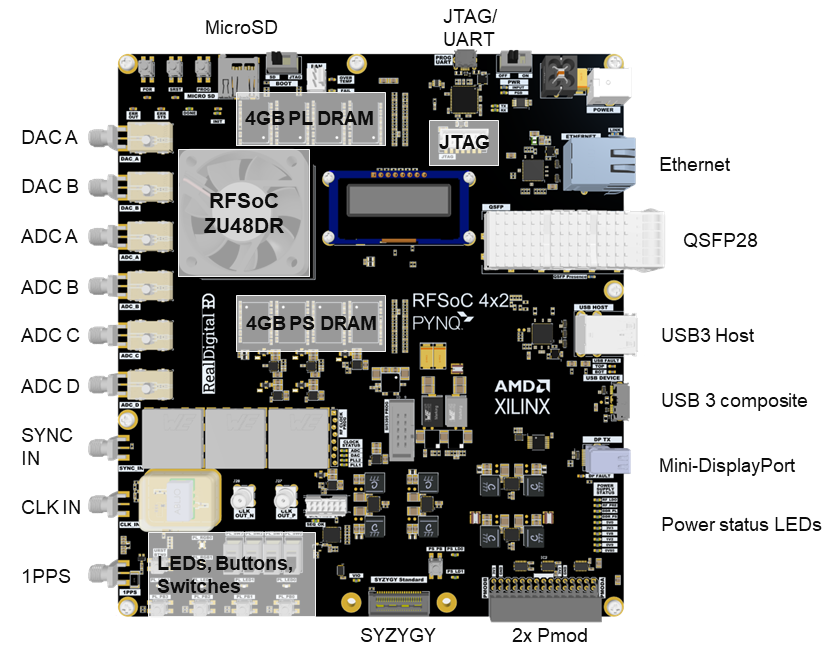}
    \caption{RFSoC 4x2 board for academic use by Real Digital [8].}
    \label{RFSoC 4*2 board}
\end{minipage}
\hfill
\begin{minipage}[b]{0.55\linewidth}
    \centering
    \includegraphics[width=\linewidth]{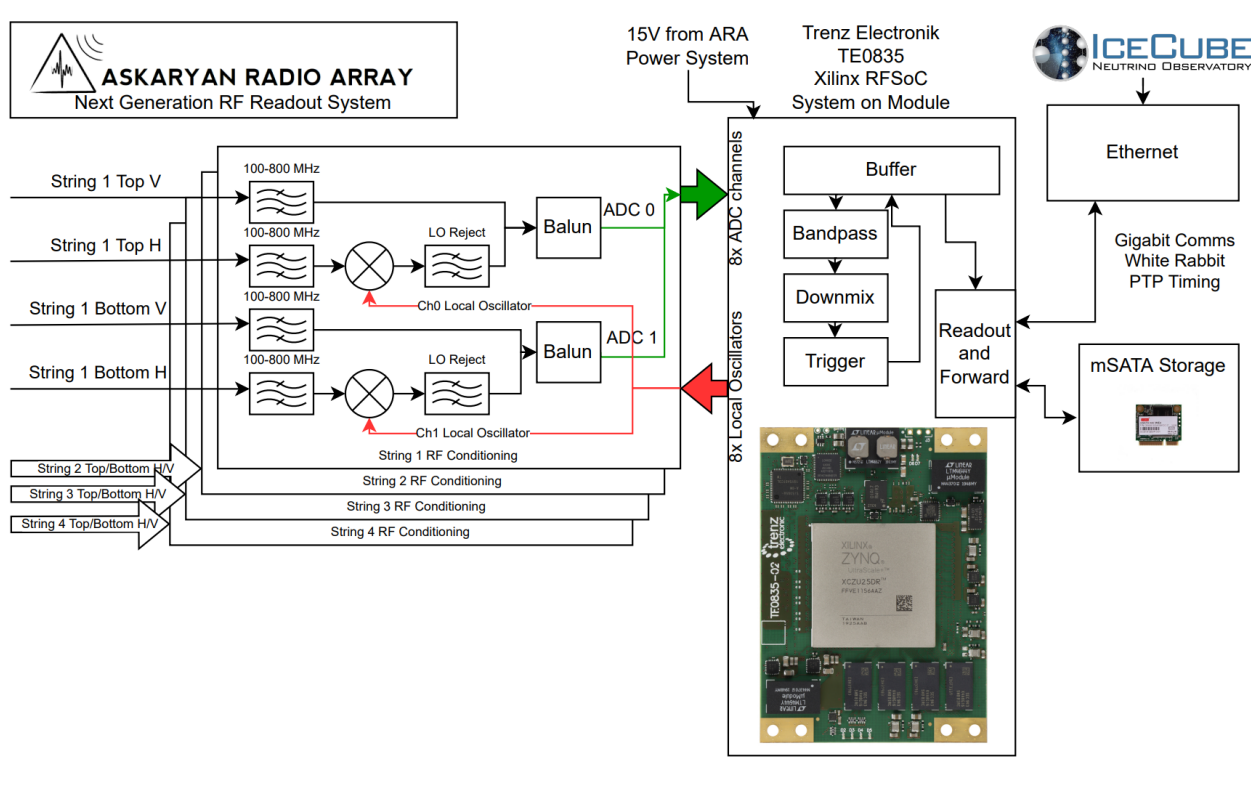}
    \caption{ARA-Next DAQ with RFSoC.}
    \label{fig:ARA-Next DAQ }
\end{minipage}
\end{figure}

\section{RFSoC for ARA-Next}\label{sec3}

The Radio Frequency System on Chip (RFSoC) is an advanced integrated circuit that combines an Field Programmable Gate Array (FPGA) with analog-to-digital converters (ADCs) and digital-to-analog converters (DACs) on a single chip. Such designs offer high-speed signal processing capabilities with minimal power consumption, making them ideal for real-time applications. The ADCs offer the high dynamic range, spectral flatness, long-term stability, and low cross-talk over their entire bandwidth \cite{ref7}. By integrating these components, RFSoC reduces latency and eliminates the need for numerous external analog parts. An example of an RFSoC board is the 4x2 RFSoC academic board by Real Digital, as shown in Figure 4. The RFSoC 4x2 board is a high-performance computing system optimized for sampling signals at up to 5 GSPS (Giga Samples Per Second) and generating signals at up to 9.85 GSPS \cite{ref8}. We are currently using this board at the University of Nebraska–Lincoln to test new trigger concepts.

RFSoC devices are exceptionally adaptable and programmable, offering a broad range of customization options. They come with development tools that simplify advanced signal processing techniques.

In neutrino detection experiments like ARA, RFSoC enhances the accuracy and efficiency of signal detection. The Askaryan effect produces radio pulses lasting only a few nanoseconds as high-energy neutrinos interact with dielectric media, requiring rapid data processing to capture these brief signals amid noise. RFSoC's advanced data converters and digital signal processing capabilities enable real-time analysis and filtering, improving the signal-to-noise ratio for a more detailed examination of such fleeting radio signals.

By noise reduction and signal analysis, RFSoC technology is expected to enable us to collect larger samples of neutrino events. The upgraded DAQ board will offer enhanced computational processing capabilities. We plan to explore and implement new triggering strategies to lower the detection threshold and increase the effective volume per station, thereby boosting ARA’s sensitivity, further solidifying ARA's position as a premier detector in the radio frequency domain.

Figure 5 illustrates the preliminary design of the RFSoC implementation for ARA-Next. The recorded signals are first passed through bandpass filters to eliminate unwanted frequencies. In the actual implementation, the RFSoC board is equipped with 8 ADC inputs, but since there are 16 total channels in a station, the number of input waveforms is reduced from 16 to 8. To achieve this, a local oscillator is used to upconvert the frequency band of one channel and combine it with another. These mixed signals are then fed into the 8 ADCs. Inside the ADCs, the original frequency bands are separated using bandpass or lowpass filters, and the remaining spectrum is downconverted using the same local oscillator set to a negative frequency. The signals from all 16 channels are then subjected to different trigger conditions. Depending on the type of trigger, signals that meet the criteria are tagged and sent to the data collection center.

\section{New Complex Trigger Strategies for ARA-Next}\label{sec4}

The success of the phased array trigger in improving detection efficiency and lowering analysis thresholds \cite{ref9} has demonstrated the potential for advanced trigger strategies in neutrino detection. This success encourages us to explore and implement unique trigger ideas within traditional ARA stations. By testing diverse trigger concepts, we aim to enhance signal processing and analysis capabilities, leading to improved detection performance and a deeper understanding of background events. We are exploring several new trigger concepts to test for ARA-Next as follow:

\begin{itemize}
\item \textbf{Trigger with 3 hits across all 16 channels (no longer limited to HPols or VPols)}

ARA has observed that signals from a VPol source several kilometers away sometimes register anomalously higher power on HPol channels, possibly due to biaxial birefringence affecting the signals' polarization \cite{ref10}. Studies on birefringent asymmetries in South Pole ice suggest that these asymmetries could help estimate the range to interaction points in in-ice neutrino interactions \cite{ref11}. Additionally, radio emissions from neutrino interactions may not be purely vertically polarized, potentially triggering either polarization antenna. Considering these factors, we plan to optimize the trigger system to detect three hits across the 16 available channels, regardless of polarization.

\item \textbf{Tag Events with a Physical Time Sequence of Hits}

The new strategy will involve tagging events based on a physical time sequence of hits to ensure that the recorded data aligns with expected physical behavior. By analyzing the timing of detected hits per channel, the system will more effectively differentiate authentic signals from random noise/pulse generated by the hardware. This approach will filter out the most common thermal background events at the trigger stage, thereby improving the reliability and quality of the collected data and prioritizing events with coherent time sequences to enhance the quality of scientific observations.

\item \textbf{Waveforms of Variable Length}

The new Data Acquisition (DAQ) system will be designed to record waveforms of varying lengths, offering increased flexibility for capturing different signal types. This feature will support improved data collection by accommodating diverse signal durations and enhancing the overall data analysis process. It will also enable fine-tuning of the trigger to detect specific signatures, such as the "double-bang" from tau neutrinos, thus improving our ability to identify and study these rare events.

\item \textbf{Utilizing Multi-Messenger Astronomy}

ARA's integration into multi-messenger astronomy will significantly broaden its observational scope by enabling the cross-referencing of data from various cosmic probes. This collaborative approach allows ARA to either inform other detectors about potential neutrino events or receive alerts from them, tagging certain events as coincident for further analysis. By aligning its findings with other astrophysical experiments, ARA can refine its detection methods, offering a more comprehensive understanding of transient cosmic phenomena. This synergy will contribute to deeper insights into ultra-high-energy particle accelerators in the distant universe.

\item \textbf{Tag or Reject Anthropogenic Events}

Effectively tagging and rejecting anthropogenic events is crucial for maintaining the accuracy of scientific data. This process involves filtering out signals like multiple pulses within a brief time frame, as shown in Figure 6, as well as prolonged radio pulses, often referred to as "fish-tails," which suggest human-made sources rather than natural origins. By rejecting anthropogenic events, we ensure that only genuine astrophysical signals are preserved for analysis.

\begin{figure}[htbp]
\centering
\begin{minipage}[b]{0.48\linewidth}
    \centering
    \includegraphics[width=\linewidth]{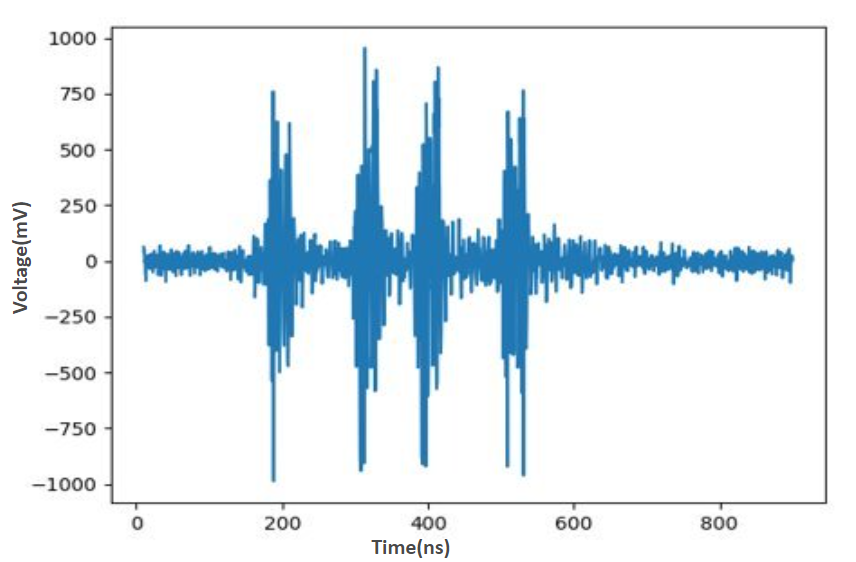}
    \caption{An example of anthropogenic event.}
    \label{fig:Anthropogenic_event}
\end{minipage}
\hfill
\begin{minipage}[b]{0.51\linewidth}
    \centering
    \includegraphics[width=0.63\linewidth]{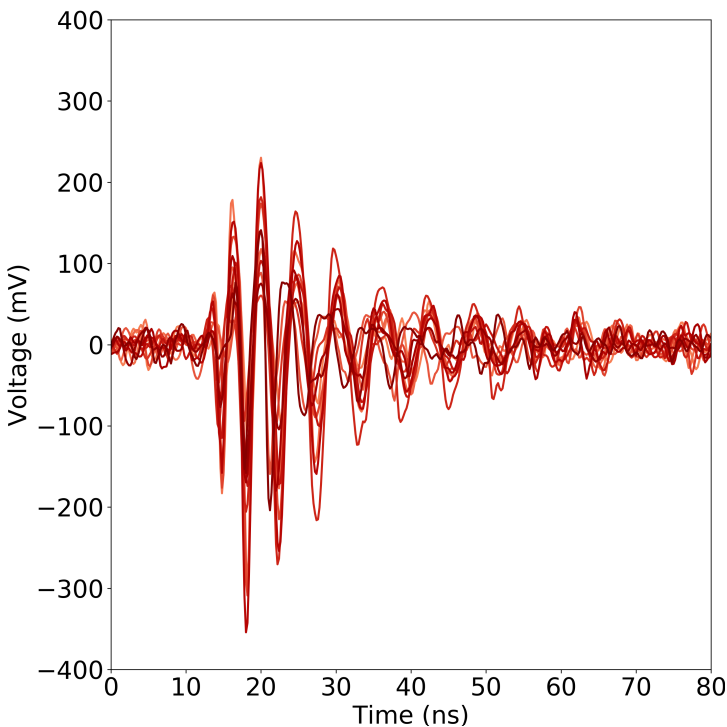} 
    \caption{Cosmic ray pulse template.}
    \label{fig:CR}
\end{minipage}
\end{figure}

\item \textbf{Tag events with possible cosmic ray pulses}

\begin{figure}
\centering
\hfill
\includegraphics[width=1\linewidth]{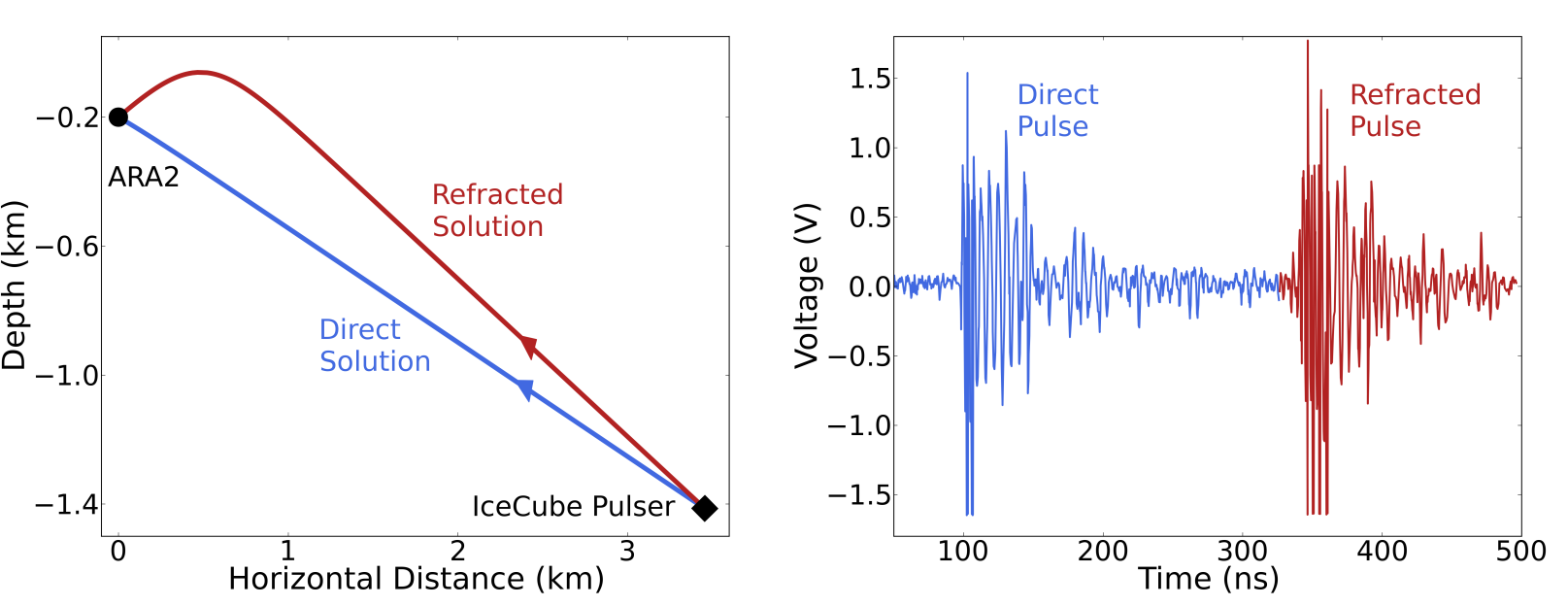}
\caption{\label{fig:DR}Double pulse event: The left plot displays the ray-tracing solution with direct (blue) and reflected (red) paths. The right plot shows the recorded two pulses on an ARA VPol antenna. }
\end{figure}

We intend to develop a trigger specifically designed to identify cosmic ray pulses by incorporating directional reconstruction at the trigger level. If the reconstructed direction points towards the ice surface, we will then perform template matching using existing cosmic ray event templates. An example of cosmic ray pulse template is shown in Figure 7. These templates provide a reference for the unique waveform patterns associated with cosmic ray interactions, enabling us to accurately recognize these high-energy events. This approach will improve our ability to detect and study cosmic ray events, ensuring that the identified waveforms correspond to known cosmic ray signatures.

\item \textbf{Tag Events with Double Pulses (n out of N Channels with Double Pulses)}

In the event of a neutrino interaction in deep ice, we expect one signal to travel directly to the antennas and another to reflect off the surface or refract and return through the ice, resulting in two distinct pulses. We plan to develop a trigger specifically designed to identify and tag such double-pulse events. This trigger will enhance our ability to detect and study significant astrophysical signals by focusing on the unique characteristics of these occurrences. An example of a double-pulse event is shown in Figure 8; it is one of the events recorded on an ARA channel during the South Pole Ice Core Experiment \cite{ref12,ref13}.

\end{itemize}

\section{Conclusion}\label{sec3}

In this proceeding, we have discussed various innovative trigger concepts for the near-future upgrade of the ARA experiment (ARA-Next). As we move forward, we are actively testing and refining these ideas to assess their viability. The incorporation of RFSoC technology will be pivotal in implementing these triggers, enabling a more flexible and sophisticated approach. Our final decision on which triggers to adopt will be guided by their effectiveness and compatibility with ARA's objectives. Under ideal conditions, the ARA community plans to replace the DAQ system with the upgraded version during the 2025-2026 pole season for at least two out of the five stations. This will not only enhance our detection capabilities but also position our new DAQ system as one of the most advanced in the radio neutrino community.

\bibliographystyle{JHEP}

\clearpage

\section*{Full Author List: ARA Collaboration (July 22, 2024)}

\noindent
S.~Ali\textsuperscript{1},
P.~Allison\textsuperscript{2},
S.~Archambault\textsuperscript{3},
J.J.~Beatty\textsuperscript{2},
D.Z.~Besson\textsuperscript{1},
A.~Bishop\textsuperscript{4},
P.~Chen\textsuperscript{5},
Y.C.~Chen\textsuperscript{5},
Y.-C.~Chen\textsuperscript{5},
B.A.~Clark\textsuperscript{6},
A.~Connolly\textsuperscript{2},
K.~Couberly\textsuperscript{1},
L.~Cremonesi\textsuperscript{7},
A.~Cummings\textsuperscript{8,9,10},
P.~Dasgupta\textsuperscript{2},
R.~Debolt\textsuperscript{2},
S.~de~Kockere\textsuperscript{11},
K.D.~de~Vries\textsuperscript{11},
C.~Deaconu\textsuperscript{12},
M.~A.~DuVernois\textsuperscript{4},
J.~Flaherty\textsuperscript{2},
E.~Friedman\textsuperscript{6},
R.~Gaior\textsuperscript{3},
P.~Giri\textsuperscript{13},
J.~Hanson\textsuperscript{14},
N.~Harty\textsuperscript{15},
K.D.~Hoffman\textsuperscript{6},
M.-H.~Huang\textsuperscript{5,16},
K.~Hughes\textsuperscript{2},
A.~Ishihara\textsuperscript{3},
A.~Karle\textsuperscript{4},
J.L.~Kelley\textsuperscript{4},
K.-C.~Kim\textsuperscript{6},
M.-C.~Kim\textsuperscript{3},
I.~Kravchenko\textsuperscript{13},
R.~Krebs\textsuperscript{8,9},
C.Y.~Kuo\textsuperscript{5},
K.~Kurusu\textsuperscript{3},
U.A.~Latif\textsuperscript{11},
C.H.~Liu\textsuperscript{13},
T.C.~Liu\textsuperscript{5,17},
W.~Luszczak\textsuperscript{2},
K.~Mase\textsuperscript{3},
M.S.~Muzio\textsuperscript{8,9,10},
J.~Nam\textsuperscript{5},
R.J.~Nichol\textsuperscript{7},
A.~Novikov\textsuperscript{15},
A.~Nozdrina\textsuperscript{1},
E.~Oberla\textsuperscript{12},
Y.~Pan\textsuperscript{15},
C.~Pfendner\textsuperscript{18},
N.~Punsuebsay\textsuperscript{15},
J.~Roth\textsuperscript{15},
A.~Salcedo-Gomez\textsuperscript{2},
D.~Seckel\textsuperscript{15},
M.F.H.~Seikh\textsuperscript{1},
Y.-S.~Shiao\textsuperscript{5,19},
S.C.~Su\textsuperscript{5},
S.~Toscano\textsuperscript{20},
J.~Torres\textsuperscript{2},
J.~Touart\textsuperscript{6},
N.~van~Eijndhoven\textsuperscript{11},
G.S.~Varner\textsuperscript{21,$\dagger$},
A.~Vieregg\textsuperscript{12},
M.-Z.~Wang\textsuperscript{5},
S.-H.~Wang\textsuperscript{5},
S.A.~Wissel\textsuperscript{8,9,10},
C.~Xie\textsuperscript{7},
S.~Yoshida\textsuperscript{3},
R.~Young\textsuperscript{1}
\\
\\
\textsuperscript{1} Dept. of Physics and Astronomy, University of Kansas, Lawrence, KS 66045\\
\textsuperscript{2} Dept. of Physics, Center for Cosmology and AstroParticle Physics, The Ohio State University, Columbus, OH 43210\\
\textsuperscript{3} Dept. of Physics, Chiba University, Chiba, Japan\\
\textsuperscript{4} Dept. of Physics, University of Wisconsin-Madison, Madison,  WI 53706\\
\textsuperscript{5} Dept. of Physics, Grad. Inst. of Astrophys., Leung Center for Cosmology and Particle Astrophysics, National Taiwan University, Taipei, Taiwan\\
\textsuperscript{6} Dept. of Physics, University of Maryland, College Park, MD 20742\\
\textsuperscript{7} Dept. of Physics and Astronomy, University College London, London, United Kingdom\\
\textsuperscript{8} Center for Multi-Messenger Astrophysics, Institute for Gravitation and the Cosmos, Pennsylvania State University, University Park, PA 16802\\
\textsuperscript{9} Dept. of Physics, Pennsylvania State University, University Park, PA 16802\\
\textsuperscript{10} Dept. of Astronomy and Astrophysics, Pennsylvania State University, University Park, PA 16802\\
\textsuperscript{11} Vrije Universiteit Brussel, Brussels, Belgium\\
\textsuperscript{12} Dept. of Physics, Enrico Fermi Institue, Kavli Institute for Cosmological Physics, University of Chicago, Chicago, IL 60637\\
\textsuperscript{13} Dept. of Physics and Astronomy, University of Nebraska, Lincoln, Nebraska 68588\\
\textsuperscript{14} Dept. Physics and Astronomy, Whittier College, Whittier, CA 90602\\
\textsuperscript{15} Dept. of Physics, University of Delaware, Newark, DE 19716\\
\textsuperscript{16} Dept. of Energy Engineering, National United University, Miaoli, Taiwan\\
\textsuperscript{17} Dept. of Applied Physics, National Pingtung University, Pingtung City, Pingtung County 900393, Taiwan\\
\textsuperscript{18} Dept. of Physics and Astronomy, Denison University, Granville, Ohio 43023\\
\textsuperscript{19} National Nano Device Laboratories, Hsinchu 300, Taiwan\\
\textsuperscript{20} Universite Libre de Bruxelles, Science Faculty CP230, B-1050 Brussels, Belgium\\
\textsuperscript{21} Dept. of Physics and Astronomy, University of Hawaii, Manoa, HI 96822\\
\textsuperscript{$\dagger$} Deceased\\

\section*{Acknowledgements}

\noindent
The ARA Collaboration is grateful to support from the National Science Foundation through Award 2013134.
The ARA Collaboration
designed, constructed, and now operates the ARA detectors. We would like to thank IceCube, and specifically the winterovers for the support in operating the
detector. Data processing and calibration, Monte Carlo
simulations of the detector and of theoretical models
and data analyses were performed by a large number
of collaboration members, who also discussed and approved the scientific results presented here. We are
thankful to Antarctic Support Contractor staff, a Leidos unit 
for field support and enabling our work on the harshest continent. We thank the National Science Foundation (NSF) Office of Polar Programs and
Physics Division for funding support. We further thank
the Taiwan National Science Councils Vanguard Program NSC 92-2628-M-002-09 and the Belgian F.R.S.-
FNRS Grant 4.4508.01 and FWO. 
K. Hughes thanks the NSF for
support through the Graduate Research Fellowship Program Award DGE-1746045. A. Connolly thanks the NSF for
Award 1806923 and 2209588, and also acknowledges the Ohio Supercomputer Center. S. A. Wissel thanks the NSF for support through CAREER Award 2033500.
A. Vieregg thanks the Sloan Foundation and the Research Corporation for Science Advancement, the Research Computing Center and the Kavli Institute for Cosmological Physics at the University of Chicago for the resources they provided. R. Nichol thanks the Leverhulme
Trust for their support. K.D. de Vries is supported by
European Research Council under the European Unions
Horizon research and innovation program (grant agreement 763 No 805486). D. Besson, I. Kravchenko, and D. Seckel thank the NSF for support through the IceCube EPSCoR Initiative (Award ID 2019597). M.S. Muzio thanks the NSF for support through the MPS-Ascend Postdoctoral Fellowship under Award 2138121. A. Bishop thanks the Belgian American Education Foundation for their Graduate Fellowship support.

\end{document}